# Interplay between out-of-plane magnetic plasmon and lattice resonance for modified resonance lineshape and near-field enhancement in double nanoparticles array[*]

Pei Ding(丁佩)[a)†], Junqiao Wang(王俊俏)[b)], Jinna He(何金娜)[b)], Chunzhen Fan(范春珍)[b)], Genwang Cai(蔡根旺)[b)], and Erjun Liang(梁二军)[b)]

[a)] Department of Mathematics and Physics, Zhengzhou Institute of Aeronautical Industry Management, Zhengzhou 450015, People's Republic of China

[b)] School of Physical Science and Engineering and Key Laboratory of Materials Physics of Ministry of Education of China, Zhengzhou University, Zhengzhou 450052, People's Republic of China

**Abstract:** Two-dimensional double nanoparticles (DNPs) arrays are demonstrated theoretically supporting the interaction of out-of-plane magnetic plasmons and in-plane lattice resonances, which can be achieved by tuning the nanoparticle height or the array period due to the height-dependent magnetic resonance and the periodicity-dependent lattice resonance. The interplay of the two plasmon modes can lead to a remarkable change in resonance lineshape and an improvement of magnetic field enhancement. Simultaneous electric field and magnetic field enhancements can be obtained in the gap regions between neighboring particles at two resonance frequencies as the interplay occurs, which present "open" cavities as electromagnetic field hot spots for potential applications on detection and sensing. The results not only offer an attractive way to tune the optical responses of plasmonic nanostructure, but also provide further insight into the plasmons interactions in periodic nanostructure or metamaterials comprising multiple elements.

[*] Project supported by the National Natural Science Foundation of China (No. 10974183, No. 11104252, No.61274012 and No. 51072184), the Specialized Research Fund for the Doctoral Program of Higher Education of China (No. 20114101110003), the Aeronautical Science Foundation of China (No. 2011ZF55015), the Basic and Frontier Technology Research Program of Henan Province (No. 112300410264 and No. 122300410162), the Foundation of University Young Key Teacher from Henan province (No. 2012GGJS-146), the Key Program of Science and Technology of Henan Education Department (No. 12A140014 and No. 13A140693) and the postdoctoral research sponsorship of Henan province (No. 2011002).

[†] Corresponding author. E-mail:dingpei@zzia.edu.cn

# 1. Introduction

Metal nanostructures with their optical responses governed by the localized surface plasmon resonances (LSPR) have drawn a great deal of attention over the past decades due to the ability of sub-wavelength confinement and intensity enhancement of the optical field [1-4]. In recent years, the coupling of two or more simple plasmonic structures has been extensively studied to achieve new and enhanced optical properties based on surface plasmons because the interaction between different plasmon modes enables spatial redistribution of the electromagnetic energy and drastic modification of the spectral characteristic [5]. One of the most studied coupling system is metallic nanoparticle (NP) dimer, such as pair of nanospheres [6], naodisks [7], nanocubes [8], nanorods [9], nanoshells [10], nanorings [11] etc.. They were shown to exhibit a large spectral shift with respect to the LSPR of individual NP and a huge enhancement of the local electric field in the gap between two NPs, being strongly dependent on the NP's composition, shape, size, relative orientation, interparticle distance and dielectric environment [9, 12-14]. Another effect of great interest is the occurrence of magnetic resonance in some specially shaped NP pairs, which are characterized by circulating displacement currents that couple strongly to the incident magnetic field [15, 16]. Intense and tunable optical magnetism at visible frequencies has potential applications in cloaking, super-resolution imaging, and ultrasensitive chemical and biological sensors [17-19].

On the other hand, periodic arrays of metallic nanostructure become one of the objects of scientific interest due to their optical responses dramatically different from those of isolated nanoparticle, especially when the grating order is transformed from radiative to evanescent [20-22]. Recently, theoretical and experimental investigations demonstrate the strong coupling between the LSPR of nanostructure and the grating diffraction (i.e. Rayleigh anomalies) in a two-dimensional array enables to excite narrow lattice plasmon resonances by suppressing radiative loss [22-26], which possess much higher quality factors $Q$ and exhibit promise applications in enhancement, directional, and polarized light emission and sensing [27, 28].

In this paper, we investigated the interaction between the in-plane lattice resonance and the out-of-plane magnetic resonance that arising from the plasmons coupling of two adjacent NPs in a two-dimensional double nanoparticles (DNPs) array. This kind of magnetic resonance in nanoparticle pairs has been reported, but the interaction with the lattice resonance was not investigated [16]. Compared to previous works that usually use the changes in in-plane architecture (lattice spacing, NP size, NP shape, interparticle distance and so on) to tune the optical response of the dimers, we indicated that the resonance wavelength of the out-of-plane magnetic plasmon can be tuned simply by changing the NP height, without changing the overall geometry of the nanostructures. By controlling the out-of-plane magnetic plasmon via NP height or tuning the in-plane lattice resonance through array period, the interaction between them can be achieved, which not only leads to a remarkable transition of the transmittance lineshape from one resonance dip toward two dips in the considered wavelength range, but also results in an improvement of magnetic field enhancement. As a result, the gaps between neighboring NPs become hot spots of electric field and magnetic field, simultaneously, at two resonance wavelengths. Different from pervious reports of achieving magnetic field enhancement through "stacked" rod pairs or disk pairs [15, 29, 30], the planar DNPs enable to provide "open" cavities for the excitation and detection of magnetic transitions in atoms and molecules, showing the potential applications of magnetic nonlinearity and magnetic sensor [17-19].

## 2. Structure and simulation method

The double nanoparticles (DNPs) array is shown schematically in Fig. 1(a), where two cylindrical gold NPs with the same shape and size are arranged in a square lattice on a silica glass. The periods in $x$ and $y$ directions are set to $p$=470 nm. The NP radius and the edge-to-edge separation between two adjacent particles are fixed at $r$=60 nm and $g$=25 nm, respectively. The NP height is varied from 60 nm to 160 nm in 10 nm steps.

Numerical electromagnetic simulations were performed in Comsol Multiphysics 4.2

with the RF module. The computational domain only contained a unit cell, as shown in Fig. 1(b), where periodic boundary conditions were employed for four lateral boundaries and perfectly matched layers (PML) were applied in the propagation direction to eliminate the nonphysical reflections at the domain boundaries. The cell was meshed with tetrahedral elements and the local mesh refinement with size less than 5 nm was applied around the NPs. A plane wave illuminated from the top of the unit with the incident direction perpendicular to the array plane and the polarization parallel to the dimer axis. The refractive index of silica glass was set to 1.45 and an experimentally measured dielectric function of gold was utilized in the simulations [31]. To obtain the transmission and extinction spectra, the power out was integrated at the bottom boundary in the wavelength range of 500−1000 nm. The transmitted power was corrected by calculating the power out from the same structure but without gold NPs in order to eliminate the interference effect. The extinction is defined as 1-$T$, with $T$ the transmittance.

## 3. Results and discussion

### 3.1. Influence of NP height on the far-field and near-field optical responses of DNPs array

Figure 2 shows the transmission spectra of DNPs array with different NP height ($h$) but fixed parameters of $p$=470 nm, $r$=60 nm and $g$=25 nm. There is only one obvious transmission dip around the wavelength of 690 nm for the arrays with smaller height ($h$<110 nm), whereas two transmission dips appear in the considered spectral range as the height is greater than 110 nm. The dip around 690 nm becomes more intense with increasing the NP height but its position is insensitive to the change of NP height, while that appearing at lower frequency shows red shift with increasing $h$, suggesting a height-dependent optical response.

To understand the transmission characteristic caused by changing the NP height, we start with the investigation of resonance origin associated with these transmission dips. For a two-dimensional (2D) nanostructure array, the lattice plasmon resonance arising from the coupling between the LSPR and the array's diffracted order should be considered. For the first-order diffraction edge, its wavelength can be calculated by $\lambda_{glass(1,0)} = d \times n_{glass}$, where $d$ is the grating constant and $n_{glass}$ is the index of glass [23]. It occurs at 681 nm using $d$=470 nm and $n_{glass}$ =1.45, as shown by the blue dashed line in Fig.2. The transmission dip occurring on the long-wavelength side of diffraction edge is a signature of coupling of light into an evanescent surface mode. Therefore, the spectral dip keeping fixed around 690 nm in the transmission spectra of DNPs array with $p$=470 nm (Fig.2) corresponds to a lattice resonance caused by coupling of dimer's LSPR and array's first diffracted order. The LSPR is located around 560nm for $h≥110$nm and associated with the longitudinal coupling of two nanoparticle's dipolar modes [6].

To give a direct insight into the transition of spectral profiles from one obvious resonance dip to two dips and understand the physics behind the height-dependent spectral features, we then investigate the near-field optical response of the DNPs arrays with different NP height, which will demonstrate the involvement of out-of-plane magnetic resonance. Figure 3 shows the averaged enhancements of electric-field (blue line) and magnetic-field (black line) in the gap between two adjacent NPs as a function of the wavelength for the array with $h$=80 nm, 100 nm, 120 nm and 140 nm, respectively. The averaged enhancement factor is defined as $\int_l |A| dl / \int_l |A_0| dl$, where $A$ and $A_0$ represent the local electric field or magnetic field at a given position with and without the DNPs array, and $l$ denotes the integration path indicated by the red dotted line in Fig. 1(b). In Fig.3, we also present the corresponding extinction spectra (red dotted line) of the DNPs arrays for comparison of the simulated far-field and near-field optical responses.

Figure 3 indicates that the DNPs array gives rise to both electric and magnetic field enhancements in the gap regions. There appears only one enhancement peak for each field occurring at different wavelength with smaller NP height ($h$=80 nm). An obvious red shift of the magnetic field enhancement peak is observed as increasing $h$ from 80nm to 100nm. When the NP height further increases ($h$>110 nm), both the electric and magnetic field enhancements split into two peaks and overlap. The electric or magnetic enhancement peak at 690 nm seems unchanged in position with the NP height while another exhibits remarkable red-shift with the further increase of $h$. The former is consistent with our above discussion about the lattice mode and the latter displays the characteristic of LSPR due to the geometry-dependence. These indicate that both electric and magnetic resonances contribute to the two transmission dips in Fig.2. The sensitivity of the magnetic field enhancement peak (or the magnetic resonant frequency) on NP height can be understood based on the capacitive coupling of two adjacent NPs. By increasing the NP height and keeping other parameters constant, the capacitance between two adjacent NPs becomes larger, which will result in the decrease of magnetic resonant frequency ($f_m \propto 1/\sqrt{LC}$)[32].

In term of the spectral line shape exhibited in Fig.3, the electric field enhancements present similar spectral feature with the optical extinctions obtained in the far field, whereas the spectrum of H-field enhancements is inconsistent with that of extinctions for smaller NP height (Fig. 3(a) or 3(b)). This is because the magnetic resonance characterized by antiphase dipole oscillation in adjacent NPs has near-zero net dipole moment (also known as dark mode [33]), therefore presents insignificant contribution to far-field spectra. Nevertheless, the averaged enhancement factor of magnetic field increases with NP height and reaches the maximum of more than 10 till $h$=120 nm and then declines as the height is further increased. Contrastively, the averaged electric field enhancement in the gap region decreases monotonously with increasing $h$ from 80 nm to 160 nm.

Figures 4 illustrates the corresponding field distributions of $E_z$, $|E|$ and $|H|$ in the cut-plane through the dimer axis for DNPs arrays with $h$=80 nm and $h$=120 nm at the peak wavelengths of averaged H-field or E-field enhancement. The field distributions of $E_z$ are provided here due to the out-of-plane electric component responsible for the out-of-plane plasmon mode. For $h$=80 nm, the out-of-plane anti-phase plasmon oscillation between two adjacent NPs is excited at $\lambda$=600 nm (Fig.4 (a)), which corresponds to the antisymmetric magnetic mode and contributes to the H-field enhancement in gap regions (Fig.4 (c) and Fig.3(a))[15, 16]. However, in-plane dipolar interaction between two particles dominates the optical response of DNPs array at $\lambda$=690 nm (the $E_x$ distribution is not shown here), leading to the E-field enhancement in the gaps (Fig. 4(e) and Fig. 3(a)). For DNPs array with smaller NPs height ($h$=80nm), the gaps between two adjacent NPs behave as hot spots of H-field or E-field at different excitation wavelength (600nm and 690nm, respectively).

For the DNPs array with $h$=120nm, the out-of-plane magnetic plasmon and the in-plane lattice resonance associated respectively with the H-field and the E-field enhancements can be excited simultaneously at both 679 nm and 711 nm (Fig. 3(c)). The two resonances carry fully the characteristics of both modes, as shown in Fig. 4(h)-4(i) and Fig. 4(k)-4(l), suggesting the near-field interaction between them. Actually, the magnetic field of the lattice surface mode has the same direction with the induced magnetic moment of the magnetic resonance (along the *y* direction). Similar to two coupled dipoles giving rise to transverse coupling, the magnetic field of the lattice resonance can interact with that of the magnetic resonance in nanoparticle pairs, leading to two mixed modes, which can be explained based on the coupled oscillators model [34]. Simultaneous E-field and H-field enhancements can be achieved in the gaps at two resonance frequencies as the interaction of both modes occurs. In addition, quadrupole components are observed in the E-field distribution (Fig. 4(h) and 4(k)) due to the excitation of multipole mode in a larger particle with higher NP [35].

### 3.2. Tuning the optical response of DNPs array through the periodicity

The interplay between out-of-plane magnetic plasmon and in-plane lattice resonance can be further explored by investigating the influence of periodicity on far-field and near-field optical properties of DNPs array with $h$=120nm. As illustrated in Fig. 5(a), there are two transmission dips for the array ($h$=120 nm, $r$=60 nm and $g$=25 nm) with smaller period and only one for larger period ($p\geq 500$ nm). The corresponding averaged H-field enhancement spectra indicate a noticeable weakening of the enhancement effect around the wavelength of 690nm, accompanied by a negligibly small enhancement peak at the long wavelength as the array period exceeds 500 nm (Fig. 5(b)). The small H-enhancement peak is expected to be contributed by the induced H-field of the electric resonance. For the case of averaged E-field enhancement, an obvious red-shift of one peak is observed as the periodicity increases (Fig. 5(c)). These phenomena can be understood very well by the fact that increasing the periodicity enables a linearly red shift of the lattice resonance and consequently non-overlapping with the magnetic resonance, which can be confirmed by investigating the averaged near-field enhancement spectra of DNPs array with $h$=80nm at different periodicity. As illustrated in Fig. 6(b), increasing the array period leads to a linearly red shift of E-field enhancement peak associated with the lattice resonance, but has negligible influence on the H-field enhancement peak caused by the out-of-plane magnetic plasmon. As a localized surface plasmon resonance (LSPR) mode, the out-of-plane magnetic resonance is strongly dependent on the geometrical structure of the DNPs rather than the lattice period. The corresponding transmission spectra of DNPs array with $h$=80nm but different periodicity are also provide in Fig. 6(a) for comparison. Therefore, the interplay of two plasmon modes due to their resonance overlapping can give rise to remarkable spectral profile and optimum H-field enhancement effect in the gap regions.

When the magnetic field of incident light is parallel to the dimer axis, no impressive spectral responses are observed because the out-of-plane magnetic resonance can't be excited under this polarization configuration, as shown by the dark line in Fig.2 for the array with $h$=120nm. In addition, we also demonstrated that a planar gold quadrumer array consisting of four identical nanocylinders arranged in a square configuration exhibits the similar optical responses as displayed in Fig.2 and Fig.3, independent of the incident polarization for symmetric configuration (not shown here). Moreover, the performance of H-field and E-field enhancement in the gap regions can be improved by further structural optimization, such as the interparticle distance, due to the magnetic and electric resonances of the nanoparticle pairs associated greatly with the coupling of two adjacent NPs.

## 4. Conclusion

In summary, we have numerically demonstrated that the two-dimensional double nanoparticles (DNPs) array supports the interplay of out-of-plane magnetic plasmon and in-plane lattice resonance. The former can be tuned more easily over a larger wavelength range by controlling the NPs height without changing the overall geometry of the nanostructure, whereas the latter depends greatly on the array period. The interaction of the two plasmon modes can give rise to a remarkable change of the resonance lineshape and result in more than 10 times averaged H-field enhancement ($h$=120nm and p=470nm) in the gap regions. Both the E-field and H-field hot spots appear simultaneously in the gaps between neighboring particles at double resonance frequencies as the interplay occurs, presenting potential applications on molecular detection and sensing. Our results not only offer an attractive way to tune the optical properties and achieve nano-localized electromagnetic field enhancements in plasmonic nanostructure, but also provide further insight into the plasmon's interactions, especially for periodic nanostructures or metamaterials comprising multiple large nanoparticles.

**List of Figure Captions:**

**Fig. 1.** (color online) (a) Schematic of a 2D square array of two cylindrical gold NPs placed on a silica glass with the incident light polarization defined. (b) Side view of a unit cell in the DNPs array, where the structural parameters of *p, r, g, h* represent the periodicity, NP radius and interparticle distance, respectively. The red line between two particles shows the integration path for calculating the averaged electromagnetic field enhancement factor.

**Fig. 2.** (color online) Transmission spectra of a series of gold cylindrical DNPs arrays with the NP height *h* varied from 80 nm (top line) to 140 nm (bottom line) in steps of 10 nm at normal-incidence with the polarization of E-field along the dimer axis. Other structural parameters are kept constant with *p*=470 nm, *r*=60 nm and *g*=25 nm. The vertical blue dashed line represents the wavelength position of the first-order diffraction edge. The dark line in the column of *h*=120 nm corresponds to the transmission spectra of the same array at normal-incidence with the polarization perpendicular to the dimer axis.

**Fig. 3.** (color online) Averaged E-field (blue line) and H-field (black line) enhancements ($\int_l |A| dl / \int_l |A_0| dl$) in the interparticle gaps as a function of wavelength for DNPs arrays with different NP height (*h*=80nm, 100nm, 120nm and 140nm). Other structural parameters are kept constant at *p*=470nm, *r*=60nm and *g*=25nm. The corresponding extinction spectra (red dotted line) of these DNPs arrays are also provided here for comparisons, whose ordinate values range from 0 to 1.

**Fig. 4.** (color online) Field distributions of $E_z$, $|E|$ and $|H|$ in the *x*-z plane for the DNPs array with different NP height at the peak wavelengths of averaged H-field or E-field enhancement. (a)-(c), *h*=80 nm and *λ*=600 nm; (d)-(f), *h*=80 nm and *λ*=690 nm; (g)-(i), *h*=120 nm and *λ*=679 nm; (j)-(l), *h*=120 nm and *λ*=711 nm.

**Fig. 5.** (color online) Transmission (a), averaged H-field (b) and E-field (c) enhancements as a function of wavelength for DNPs arrays with different periodicity of *p* and fixed structural parameters of *h*=120, *r*=60nm and *g*=25nm.

**Fig.6.** (color online) Transmission (a), averaged H-field and E-field (b) enhancements as a function of wavelength for DNPs arrays with different periodicity of *p* and fixed structural parameters of *h*=80, *r*=60nm and *g*=25nm.

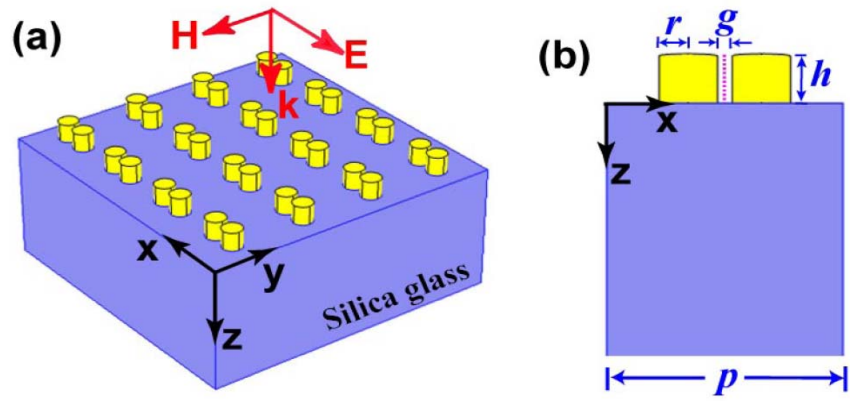

**Figure 1**

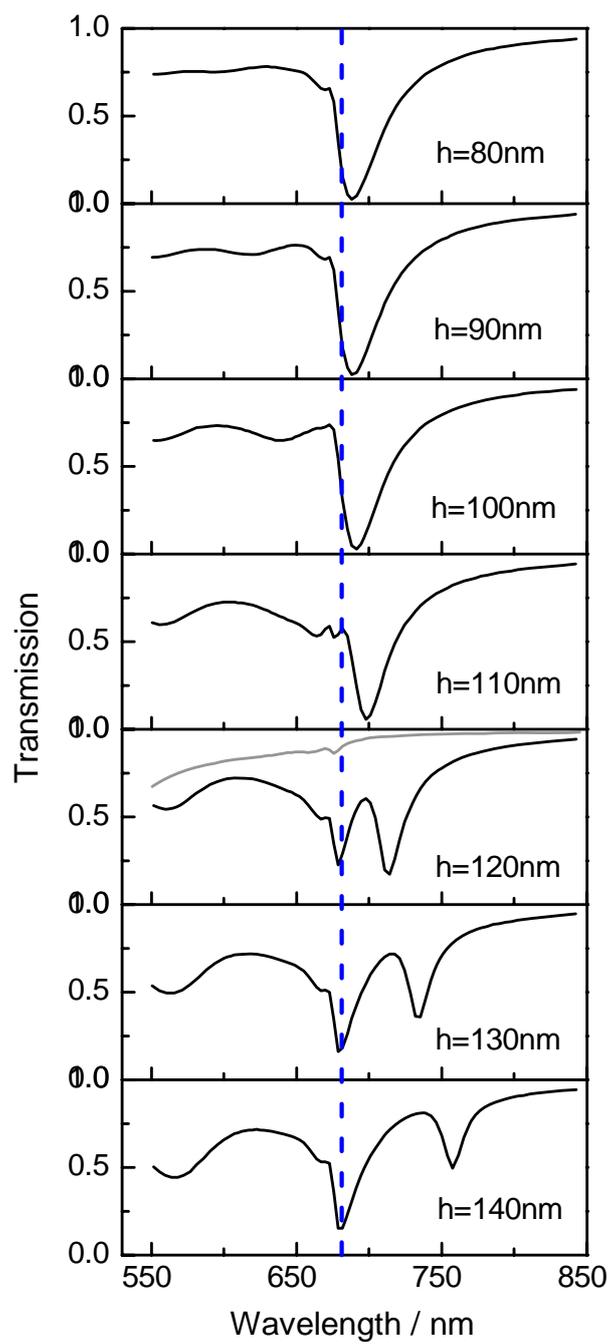

**Figure 2**

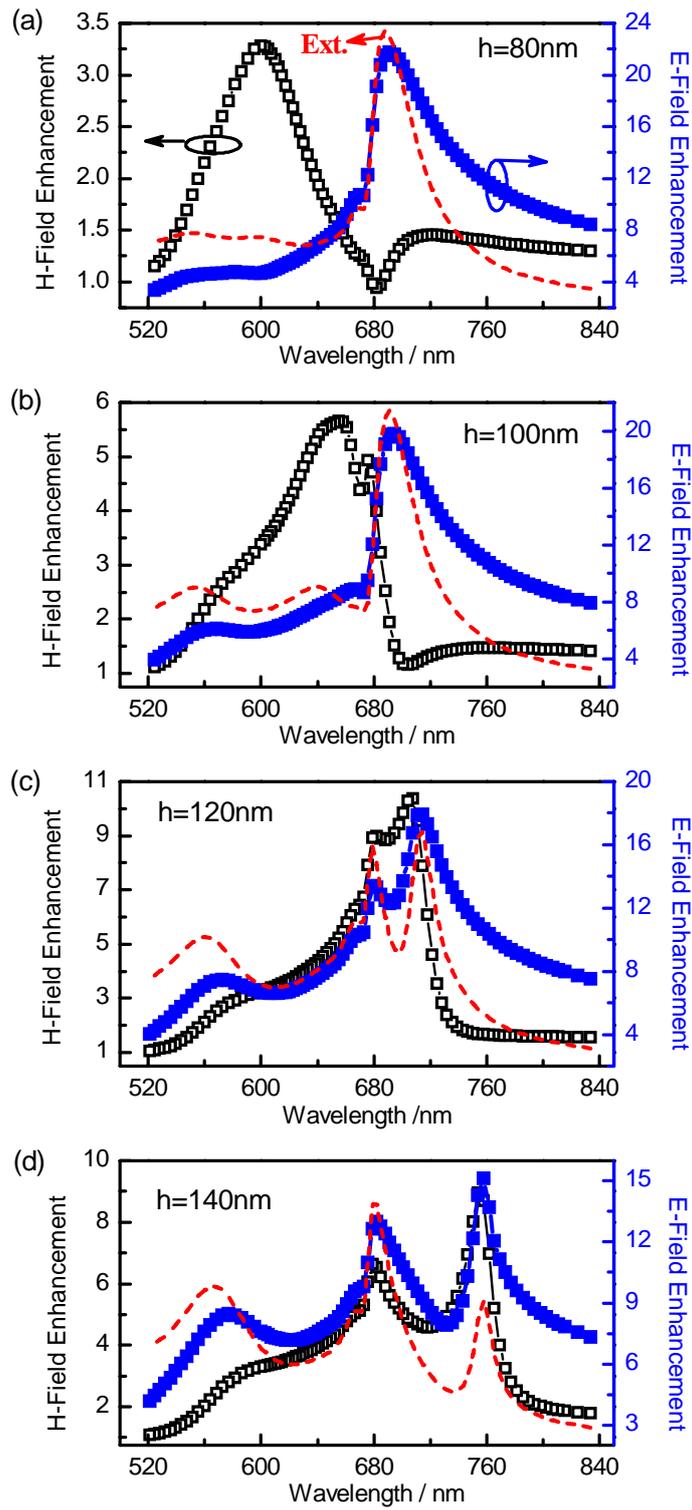

**Figure 3**

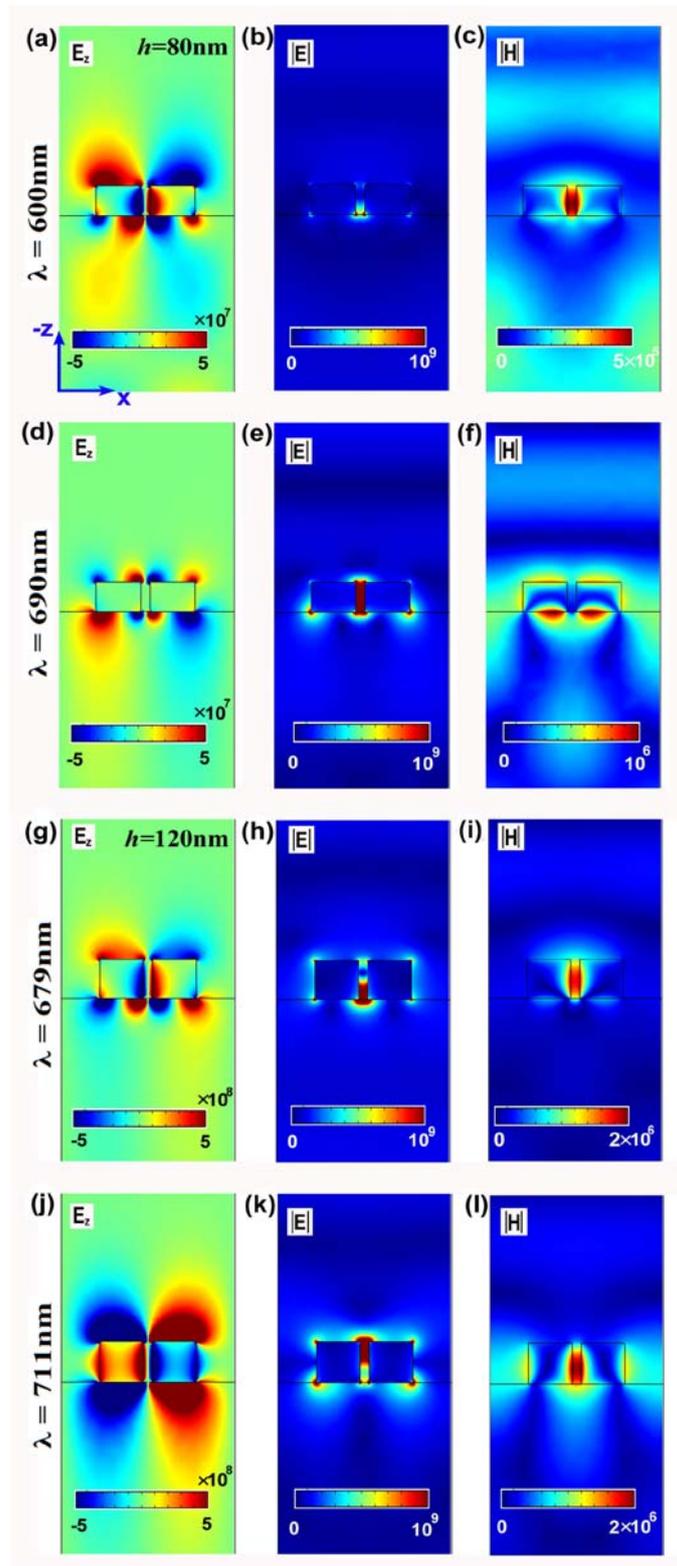

**Figure 4**

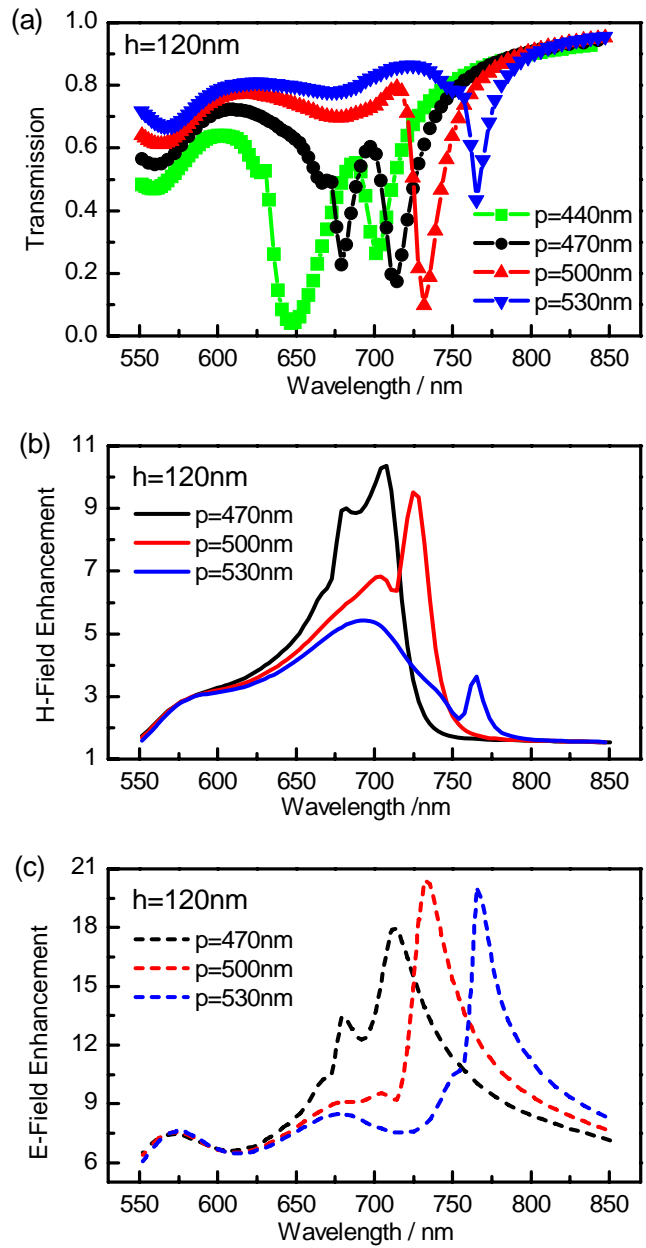

**Figure 5**

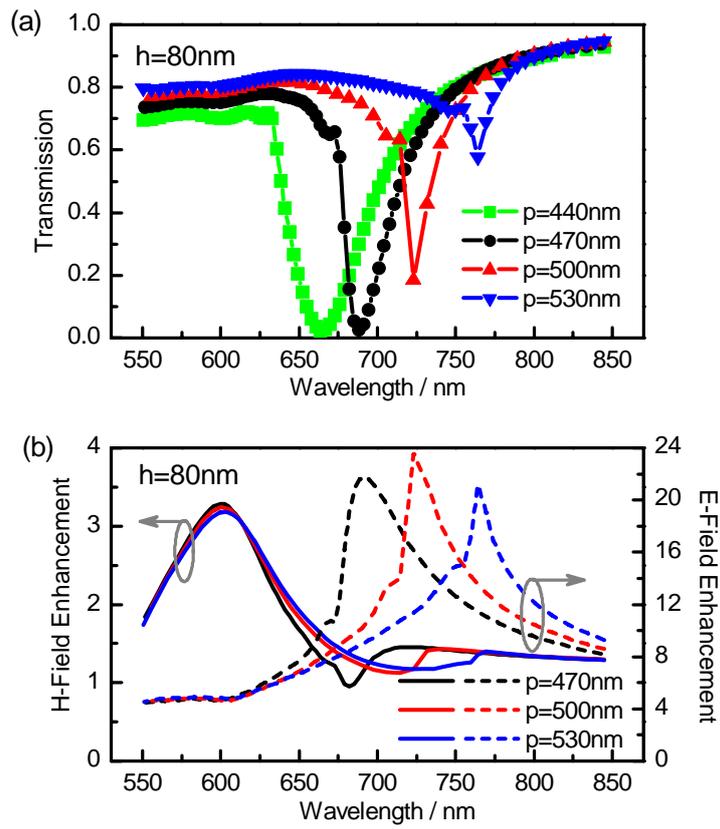

**Figure 6**